\documentclass[10pt]{aastex63}
\usepackage{graphicx}
\usepackage{epstopdf}
\usepackage{color}

\newcommand{\D}[1]{\, d #1 \,}
\newcommand{\vc}[1]{{\bf#1}}
\newcommand{\tens}[1]{{\bf#1}}
\newcommand{\vch}[1]{{\bf \hat{#1}}}
\newcommand{\avg}[1]{\left< #1 \right>}
\newcommand{\ft}[1]{\widetilde{#1}}
\newcommand{\mcl}[1]{\mathcal{#1}}
\newcommand{\eps}{\varepsilon}
\newcommand{\vphi}{\varphi}
\newcommand{\Bo}{B_{0\bot}}
\newcommand{\phis}{\vphi_s}
\newcommand{\e}[2]{\hat{e}_{{#1}_s{#2}}}
\newcommand{\sqrtd}[1]{\sqrt{d\bf{#1}} \:}

\newcommand{\sqrtdk}[2]{\sqrt{d\vc{k}#1_{#2}} \:}
\newcommand{\expk}[2]{e^{i\vc{k}#1_{#2}\vc{r}}}
\newcommand{\Fk}[2]{F({\vc{k}#1_{#2}})}
\newcommand{\exi}[3]{\hat{e}_{#1_si#2_#3}}
\newcommand{\Tijk}[2]{T_{i#1_#2 j#1_#2}(\vch{k}#1_#2)}
\newcommand{\ksijk}[2]{\xi_{j#1_#2}(\vc{k}#1_#2)}
\newcommand{\kk}[2]{\vc{k}#1_#2}

\definecolor{dkyellow}{rgb}{0.85,0.85,0}

\begin{document}

\title{Identification of MHD compressible modes in interstellar plasma with synchrotron emission polarization}


\author{Alexey Chepurnov}
\affiliation{Deutsches Elektronen-Synchrotron DESY, Zeuthen, Germany}


\author{Reinaldo Santos de Lima}
\affiliation{University of S\~{a}o Paulo, Brazil}

\author{Sarah Appleby}
\affiliation{Edinburgh University, UK}


\begin{abstract}
We provide a procedure for identification of dominating compressible and Alfv\'enic MHD modes or isotropic turbulence in synchrotron emission polarization maps of Galactic objects. The results for the region of North Galactic Pole, Orion molecular cloud complex and the star-forming complex Cygnus X are presented.
\end{abstract} 

\keywords{methods: data analysis --- turbulence --- ISM: magnetic fields --- techniques: miscellaneous}


\section{Introduction} \label{sect:intro}

Turbulence is ubiquitous and plays crucial roles in various interstellar processes including star formation and dynamo and transport processes. Interstellar medium (ISM) is magnetized, indicating the magneto-hydrodynamic (MHD) nature of turbulence in ISM. As the result, it is theoretically expected that interstellar turbulence can have three MHD modes, Alfv\'en, fast and slow compressible modes \citep[]{CL03, YL04}. It is challenging to remotely diagnose the MHD modes of turbulence except for the nearby solar wind turbulence which can be directly detected by space probes, \citep{E02}. Here we report the detection of MHD modes of  interstellar turbulence using Galactic synchrotron polarization data.

We developed a new technique based on polarization data of synchrotron emission which does not require estimation of power spectrum from observational data. 

Turbulence anisotropy is imprinted in the set of Stokes parameters (I, Q, U), which characterize the polarization state of radiation.  
We use here the values (I+Q)/2 and U/2, whose emissivities are proportional to the squared picture-plane projection of the magnetic field, and the product of perpendicular picture plane projections respectively. 

Their variances as functions of Stokes parameters positional angle (hereafter we name them "signatures") can be parametrized so that one of the parameters can be used to identify the dominating MHD mode.


\section{Magnetic field}  \label{sect:mf}

Let us consider a model where the turbulent magnetic field is a homogeneous random field which can be described with its spectral representation (see Sect. \ref{sect:sr} for the details):
\begin{equation} \label{eq:B}
B_i(\vc{r}) = \int \sqrtd{k} e^{i\vc{k}\vc{r}} F(\vc{k}) \,  T_{ij}(\vch{k}) \xi_j(\vc{k}) 
\end{equation}
\noindent where the spectral tensor $T_{ij}$ and random field $\xi$ conform the following rules: $T_{ij} = T_{il}T_{jl}$, $T_{ij}=T_{ji}$, $\avg{\xi_i(\vc{k})\xi_j^*(\vc{k}')} = \delta_{ij} \delta_{\vc{k}\vc{k}'}$, $\xi_i(-\vc{k}) = \xi_i^*(\vc{k})$ and $F(\vc{k})$ is the square root of the scalar part of the power spectrum.

\begin{figure}[h]
\begin{center}
\includegraphics[scale=0.7]{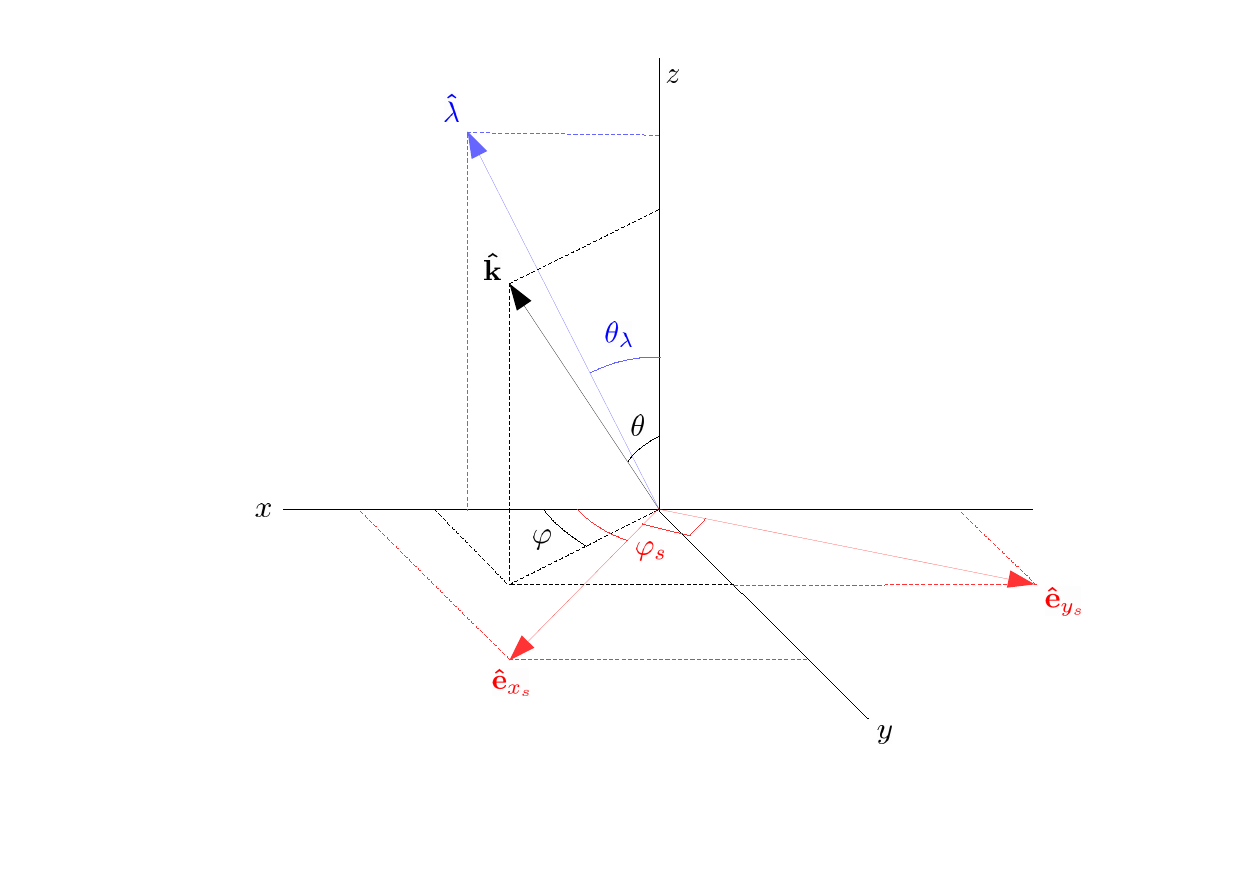}
\end{center}
\caption{Vectors and angles involved in calculating of spectral tensor projections. $\vch{\lambda}$ is the symmetry axis, $\vch{k}$ is the direction of the wavevector, vectors $\vch{e}_{x_s}$ and $\vch{e}_{y_s}$ define the frame of the Stokes parameters, the axis $Oz$ points to the observer.\label{fig:vectors}} 
\end{figure}

The terms forming the spectral representation are different for different modes. We assume here that the turbulent magnetic field has axial symmetry, defined by the mean magnetic field with the direction $\vch{\lambda}$, see Fig. \ref{fig:vectors}. 

Let us define magnetic field power spectra for different MHD modes, following \cite{LP12}.

The spectral tensor of the axially-symmetrical magnetic field can be combined of the two parts, $T_E$ and $T_F$:
\begin{equation} \label{eq:TE}
T_{E,ij}(\vch{k}) = \delta_{ij}-\hat{k}_i\hat{k}_j
\end{equation}
\begin{equation} \label{eq:TF}
T_{F,ij}(\vch{k}) = \frac{
(\vch{k}\vch{\lambda})^2\hat{k}_i\hat{k}_j+\hat{\lambda}_i\hat{\lambda}_j-(\vch{k}\vch{\lambda})(\hat{k}_i\hat{\lambda}_j+\hat{\lambda}_i\hat{k}_j)
}{1-(\vch{k}\vch{\lambda})^2}
\end{equation}

Then, the terms defining power spectra of the MHD modes are given below.

\noindent Alfv\'enic mode:
\begin{equation} \label{eq:alf_T}
T_{ij}(\vch{k}) = T_{E,ij}(\vch{k})-T_{F,ij}(\vch{k})
\end{equation}
\begin{equation} \label{eq:alf_F0}
 F^2_0(\vch{k}) = \exp \left( -M_a^{-4/3}\frac{|\vch{k}\vch{\lambda}|}{(1-(\vch{k}\vch{\lambda})^2)^{1/3}} \right)
\end{equation}
Fast mode:
\begin{equation} \label{eq:fast_T}
T_{ij}(\vch{k}) = T_{F,ij}(\vch{k})
\end{equation}
\begin{equation} \label{eq:fast_F0}
 F^2_0(\vch{k}) =
  \left\{ 
    \begin{array}{ll} 
      1, & \beta \ll 1 \\ 
      1-(\vch{k}\vch{\lambda})^2, & \beta \gtrsim 1
    \end{array} 
  \right. 
\end{equation}
Slow mode (high-$\beta$ only):
\begin{equation} \label{eq:slow_T}
T_{ij}(\vch{k}) = T_{F,ij}(\vch{k})
\end{equation}
\begin{equation} \label{eq:slow_F0}
 F^2_0(\vch{k}) = \exp \left( -M_a^{-4/3}\frac{|\vch{k}\vch{\lambda}|}{(1-(\vch{k}\vch{\lambda})^2)^{1/3}} \right)
\end{equation}
\noindent here $M_a$ is the Alfv\'enic Mach number and $ F^2_0$ is the factor defining the anisotropy of the scalar part of the power spectrum $F^2$: 
\begin{equation} \label{eq:F}
 F^2(\vc{k}) \propto \frac{e^{-\left(\frac{2\pi}{kL}\right)^2}}{k^{11/3}} F^2_0(\vch{k})
\end{equation}
\noindent where $L$ is the injection scale.

We are interested in calculating of quantities related to Stokes parameters, which are expressed via magnetic field in the following way (omitting the scaling coefficient):
\begin{equation} \label{eq:i}
 I(\vc{R}) = \int w(z) \D{z} (B^2_{x_s}(\vc{r})+B^2_{y_s}(\vc{r}))
\end{equation}
\begin{equation} \label{eq:q}
 Q(\vc{R}) = \int w(z) \D{z} (B^2_{x_s}(\vc{r})-B^2_{y_s}(\vc{r}))
\end{equation}
\begin{equation} \label{eq:u}
 U(\vc{R}) = \int w(z) \D{z} 2B_{x_s}(\vc{r})B_{y_s}(\vc{r})
\end{equation}
\noindent here $w$ is the window function defining the borders of the emitting structure. We assume here that the lines of sight corresponding to picture plane are parallel and the 2D vector $\vc{R}$ gives us the coordinates in the picture plane. In some cases it is not completely true, but we assume that it is an acceptable approximation.

In this work we often use numerical evaluation of Eq. \ref{eq:B} to generate magnetic field for a particular MHD mode and calculate correspondent Stokes parameter maps by applying\footnote{With addition of a constant component of magnetic field.} Eqs. \ref{eq:i} - \ref{eq:u}. Hereafter we name such data synthetic.


\section{Emissivity correlation functions and power spectra} \label{sect:ecf}

In this section we shall derive correlation functions and power spectra of emissivities related to MHD mode signatures.


\subsection{The case of (I+Q)/2}

The corresponding emissivity can be defined as follows:
\begin{equation} \label{eq:exx}
\begin{array}{ll}
\eps_{xx}(\vc{r}) 
  & = (\Bo\cos\phis+B_i \e{x}{i})^2 \\
  & = \Bo^2\cos^2\phis + 2\Bo\cos\phis B_i \e{x}{i} + (B_i  \e{x}{i})^2
\end{array}
\end{equation}
where $\Bo$ is the picture-plane projection of mean magnetic field.

Using the spectral representation of turbulent magnetic field Eq. \ref{eq:B}, we can write the following expression for the emissivity correlation function:
\begin{equation} \label{eq:Cxx}
\begin{array}{lll}
C_{xx}(\vc{r})
  & \equiv & \avg{\eps_{xx}(\vc{r})\eps_{xx}(\vc{0})} \\ 
  & = & \left<\left(\Bo^2\cos^2\phis 
    + 2\Bo\cos\phis      \int \sqrtdk{}{1} \expk{}{1} \Fk{}{1} \, \exi{x}{}{1} \Tijk{}{1} \ksijk{}{1} \right.\right. \\
  & + &                  \int \sqrtdk{}{2} \expk{}{2} \Fk{}{2} \, \exi{x}{}{2} \Tijk{}{2} \ksijk{}{2} \\
  & \; \cdotp & \left.\! \int \sqrtdk{}{3} \expk{}{3} \Fk{}{3} \, \exi{x}{}{3} \Tijk{}{3} \ksijk{}{3} \right) \\
  & \; \cdotp & \left(\Bo^2\cos^2\phis 
    + 2\Bo\cos\phis               \int \sqrtdk{'}{1} \Fk{'}{1} \, \exi{x}{'}{1} \Tijk{'}{1} \ksijk{'}{1} \right. \\
  & + &                           \int \sqrtdk{'}{2} \Fk{'}{2} \, \exi{x}{'}{2} \Tijk{'}{2} \ksijk{'}{2} \\
  & \; \cdotp & \left.\left. \!\! \int \sqrtdk{'}{3} \Fk{'}{3} \, \exi{x}{'}{3} \Tijk{'}{3} \ksijk{'}{3} \right) \right> \\
\end{array}
\end{equation}
The combinations of wavevectors, giving non-zero contribution to Eq. \ref{eq:Cxx}, can be found accounting for Eqs. \ref{eq:vxisym} and \ref{eq:vxicorr}. The correspondent list is presented in Tab.\ref{tab:xx}.
\begin{deluxetable}{lll}[h]
\tablecaption{List of wavevector combinations, giving non-zero contribution to Eq. \ref{eq:Cxx} (the correlation function of emissivity of (I+Q)/2). \label{tab:xx}}
\tablehead{
  \colhead{combinations of $\xi$} & \colhead{combinations of $\vc{k}$} & \colhead{note}
}
\startdata
$\avg{\ksijk{}{1}\ksijk{'}{1}}$  & $\kk{}{1} = -\kk{'}{1}$ & \, \\ 
$\avg{\ksijk{}{2}\ksijk{'}{2}\ksijk{}{3}\ksijk{'}{3}}$  & \, & \, \\
\, & $\kk{}{2} = -\kk{'}{2} \; \& \; \kk{}{3} = -\kk{'}{3}$ & \, \\
\, & $\kk{}{2} = -\kk{'}{3} \; \& \; \kk{}{3} = -\kk{'}{2}$ & \, \\
\, & $\kk{}{2} = -\kk{}{3} \; \& \; \kk{'}{2} = -\kk{'}{3}$ & const \\
$\avg{\ksijk{}{2}\ksijk{}{3}}$  & $\kk{}{2} = -\kk{}{3}$ & const \\ 
$\avg{\ksijk{'}{2}\ksijk{'}{3}}$  & $\kk{'}{2} = -\kk{'}{3}$ & const \\ 
\enddata
\tablecomments{
Combinations giving constant contribution will be omitted.
}
\end{deluxetable}

Evaluating averaging in Eq. \ref{eq:Cxx} and accounting for Tab.\ref{tab:xx} and Eq. \ref{eq:vt}, we have:
\begin{equation} \label{eq:cxxfin}
\begin{array}{lll}
C_{xx}(\vc{r})
  & = & (2\Bo\cos\phis)^2 \int \D{\vc{k}} F^2(\vc{k}) e^{i\vc{k}\vc{r}} \hat{e}_{x_si} \hat{e}_{x_sj} T_{ij}(\vch{k}) \\
  & + & 2\left( \int \D{\vc{k}} F^2(\vc{k}) e^{i\vc{k}\vc{r}} \hat{e}_{x_si} \hat{e}_{x_sj} T_{ij}(\vch{k}) \right)^2 \\
\end{array}
\end{equation}
Then the correspondent power spectrum and its components can be written as follows:
\begin{equation} \label{eq:fxx}
F^2_{xx}(\vc{k}) = F^2_{xx, l}(\vc{k})+F^2_{xx, q}(\vc{k}) 
\end{equation}
\noindent where
\begin{equation} \label{eq:fxxlin}
F^2_{xx, l}(\vc{k}) = (2\Bo\cos\phis)^2 \mcl{F}^2_{x_sx_s}(\vc{k})
\end{equation}
\begin{equation} \label{eq:fxxquad}
F^2_{xx, q}(\vc{k}) = 2 \int \D{\vc{k}'} \mcl{F}^2_{x_sx_s}(\vc{k'})\mcl{F}^2_{x_sx_s}(\vc{k} - \vc{k'})
\end{equation}
\noindent and
\begin{equation} \label{eq:fxxbas}
\mcl{F}^2_{x_sx_s}(\vc{k}) = F^2(\vc{k}) \hat{e}_{x_si} \hat{e}_{x_sj} T_{ij}(\vch{k}) 
\end{equation}
Hereafter $F^2_{xx, l}$ and $F^2_{xx, q}$ will be referred as linear and quadratic terms of the emissivity power spectrum of (I+Q)/2.


\subsection{The case of U/2}

The corresponding emissivity can be defined as follows:
\begin{equation} \label{eq:exy}
\begin{array}{ll}
\eps_{xy}(\vc{r}) 
  & = (\Bo\cos\phis+B_i \e{x}{i})(-\Bo\sin\phis+B_i \e{y}{i}) \\
  & = -\Bo^2\cos\phis\sin\phis - \Bo\sin\phis B_i \e{x}{i} + \Bo\cos\phis B_i \e{y}{i} + B_i\e{x}{i} B_{i'}\e{y}{i'}
\end{array}
\end{equation}

In a similar way we can obtain the following expressions for its power spectrum and its components:
\begin{equation} \label{eq:fxy}
F^2_{xy}(\vc{k}) = F^2_{xy, l}(\vc{k})+F^2_{xy, q}(\vc{k}) 
\end{equation}
\noindent where
\begin{equation} \label{eq:fxylin}
F^2_{xy, l}(\vc{k}) = (\Bo\sin\phis)^2 \mcl{F}^2_{x_sx_s}(\vc{k}) - 2\Bo^2\sin\phis\cos\phis \mcl{F}^2_{x_sy_s}(\vc{k}) + (\Bo\cos\phis)^2 \mcl{F}^2_{y_sy_s}(\vc{k})
\end{equation}
\begin{equation} \label{eq:fxyquad}
F^2_{xy, q}(\vc{k}) =  \int \D{\vc{k}'} \mcl{F}^2_{x_sx_s}(\vc{k'})\mcl{F}^2_{y_sy_s}(\vc{k} - \vc{k'}) + \int \D{\vc{k}'} \mcl{F}^2_{x_sy_s}(\vc{k'})\mcl{F}^2_{x_sy_s}(\vc{k} - \vc{k'})
\end{equation}
\noindent and
\begin{equation} \label{eq:fxxbas1}
\mcl{F}^2_{x_sx_s}(\vc{k}) = F^2(\vc{k}) \hat{e}_{x_si} \hat{e}_{x_sj} T_{ij}(\vch{k}) 
\end{equation}
\begin{equation} \label{eq:fxybas1}
\mcl{F}^2_{x_sy_s}(\vc{k}) = F^2(\vc{k}) \hat{e}_{x_si} \hat{e}_{y_sj} T_{ij}(\vch{k}) 
\end{equation}
\begin{equation} \label{eq:fyybas1}
\mcl{F}^2_{y_sy_s}(\vc{k}) = F^2(\vc{k}) \hat{e}_{y_si} \hat{e}_{y_sj} T_{ij}(\vch{k}) 
\end{equation}
Here $F^2_{xy, l}$ and $F^2_{xy, q}$ are linear and quadratic terms of the emissivity power spectrum of U/2.


\section{The mode signature} \label{sect:sig}

If the emissivity has 3D power spectrum $F^2(\vc{k})$ and it is projected as described in Sect. \ref{sect:mf}, the correspondent picture plane power spectrum can be written as follows:
\begin{equation} \label{eq:Phi}
\Phi^2(\vc{K}) = (2\pi)^2 \int_{-\infty}^\infty |\tilde{w}(k_z)|^2 F^2(\vc{k}) \D{k_z}
\end{equation}
\noindent where $\vc{K} \equiv (k_x,k_y)$ and $\tilde{w}$ is the Fourier-transformed window function.

Then we can write the following expression for related signature:
\begin{equation} \label{eq:s}
\begin{array}{ll}
s(\phis) & = \int \D{\vc{K}} \Phi^2(\vc{K}) f^2(K) \\
         & = \int w_{fz}(\vc{k}) \D{\vc{k}}  F^2(\vc{k})
\end{array}
\end{equation}
\noindent where $f^2(K)$ is the filter which includes the beam smoothing and high-pass filter for improving statistics\footnote{This filter is also used for mode identification, see below.} with the filtering scale $L_f$. The spectral window function $w_{fz}$ is defined as follows:
\begin{equation} \label{eq:wfz}
w_{fz}(\vc{k}) = (2\pi)^2 |\tilde{w}(k_z)|^2 f^2(K)
\end{equation}

According to Eqs. \ref{eq:fxx} and \ref{eq:fxy}, "theoretical" signatures consist of linear and quadratic terms\footnote{For observed signatures odd powers of turbulent magnetic field are relevant too, because ensemble averaging is replaced with the spatial one in this case (see Sect. \ref{sect:odd}).}. The linear term is important because we expect it to dominate in the signal. It is also important that we can obtain analytic results in this case. We shall consider it in the next section.


\subsection{Signature linear term} \label{sect:sig_lin}

To calculate the signatures $s_{xx}$ and $s_{xy}$ we have to substitute Eq. \ref{eq:fxxlin} and Eq. \ref{eq:fxylin} to Eq. \ref{eq:s} (we omit here the common factor $\Bo^2$):
\begin{equation} \label{eq:sxx_orig}
s_{xx}(\phis) = (2\cos\phis)^2 \int w_{fz}(\vc{k}) \D{\vc{k}} F^2(\vc{k}) T_{x_sx_s}(\vch{k})
\end{equation}
\begin{equation} \label{eq:sxy_orig}
s_{xy}(\phis) = \int w_{fz}(\vc{k}) \D{\vc{k}} F^2(\vc{k}) \left(\sin^2\phis\; T_{x_sx_s}(\vch{k}) - 2\sin\phis\cos\phis T_{x_sy_s}(\vch{k}) + \cos^2\phis\; T_{y_sy_s}(\vch{k}) \right)
\end{equation}
\noindent where correspondent tensor projections are as follows:
\begin{equation} \label{eq:Txx}
T_{x_sx_s}(\vch{k}) = \hat{e}_{x_si} \hat{e}_{x_sj} T_{ij}(\vch{k}) 
\end{equation}
\begin{equation} \label{eq:Txy}
T_{x_sy_s}(\vch{k}) = \hat{e}_{x_si} \hat{e}_{y_sj} T_{ij}(\vch{k}) 
\end{equation}
\begin{equation} \label{eq:Tyy}
T_{y_sy_s}(\vch{k}) = \hat{e}_{y_si} \hat{e}_{y_sj} T_{ij}(\vch{k}) 
\end{equation}

Let us begin with the signature $s_{xx}$. We can use spectral tensor geometrical properties, described in Sect. \ref{sect:geom}, to calculate correspondent projection $\vch{e}_{x_s} \tens{T}_\alpha \vch{e}_{x_s}$ (here $\alpha \in \{c,a\}$ is a tensor type, compressible or Alfv\'enic). 

As $\tens{T}_\alpha$ is an orthogonal projector having 1D range, the rightmost multiplication gives us the projection of $\vch{e}_{x_s}$ to the line defined by the range basis vector $\vch{e}_\alpha$, see Eq. \ref{eq:proj}. The remaining multiplication gives us the scalar product of this projection with $\vch{e}_{x_s}$. Therefore we have:
\begin{equation} \label{eq:proj_xx}
\begin{array}{ll}
\vch{e}_{x_s} \tens{T}_\alpha \vch{e}_{x_s} 
  & = (\vch{e}_\alpha \vch{e}_{x_s})^2 \\
  & = (\hat{e}_{\alpha,x} \cos \phis + \hat{e}_{\alpha,y} \sin \phis)^2 \\
  & = (\hat{e}_{\alpha,y}^2 - \hat{e}_{\alpha,x}^2) \sin^2 \phis + \hat{e}_{\alpha,x}^2 
    + 2 \hat{e}_{\alpha,x} \hat{e}_{\alpha,y} \cos \phis \sin \phis\\
\end{array}
\end{equation}

After substitution of Eq. \ref{eq:proj_xx} into Eq. \ref{eq:sxx_orig} we have the following expression for $s_{xx}$:
\begin{equation} \label{eq:sxx}
s_{xx}(\phis) = (a_{xx}\sin^2\phis+b_{xx}+c_{xx}\sin 2\phis)\cos^2\phis
\end{equation}
where
\begin{equation} \label{eq:axx}
a_{xx} = 4 \left(\avg{\hat{e}_{\alpha,y}^2}_\Omega - \avg{\hat{e}_{\alpha,x}^2}_\Omega\right)  
\end{equation}
\begin{equation} \label{eq:bxx}
b_{xx} = 4 \avg{\hat{e}_{\alpha,x}^2}_\Omega  
\end{equation}
\begin{equation} \label{eq:cxx}
c_{xx} = 4 \avg{\hat{e}_{\alpha,x} \, \hat{e}_{\alpha,y}}_\Omega  
\end{equation}
and $\avg{...}_\Omega$ denotes integration over wavevector directions with the proper weight:
\begin{equation} \label{eq:iomega}
\avg{h}_\Omega \equiv \int w_\Omega(\vch{k}) \D{\Omega} h(\vch{k})
\end{equation}
where
\begin{equation} \label{eq:womega}
w_\Omega(\vch{k}) = \int_0^\infty k^2\D{k} w_{fz}(k\vch{k}) F^2(k\vch{k})
\end{equation}

In a similar way we can obtain the following expression for $s_{xy}$:
\begin{equation} \label{eq:sxy}
s_{xy}(\phis) = a_{xy}\cos 4\phis + b_{xy} + c_{xy}\sin 4\phis
\end{equation}
where
\begin{equation} \label{eq:axy}
a_{xy} = \frac{1}{2} \left(\avg{\hat{e}_{\alpha,y}^2}_\Omega - \avg{\hat{e}_{\alpha,x}^2}_\Omega\right)  
\end{equation}
\begin{equation} \label{eq:bxy}
b_{xy} = \frac{1}{2} \left(\avg{\hat{e}_{\alpha,y}^2}_\Omega + \avg{\hat{e}_{\alpha,x}^2}_\Omega\right)  
\end{equation}
\begin{equation} \label{eq:cxy}
c_{xy} = -\avg{\hat{e}_{\alpha,x} \, \hat{e}_{\alpha,y}}_\Omega  
\end{equation}

Then we can find the following relations between the parameters of $xx$- and $xy$-signatures:
\begin{equation} \label{eq:axxxy}
a_{xy} = \frac{1}{8} a_{xx}
\end{equation}
\begin{equation} \label{eq:bxxxy}
b_{xy} = \frac{1}{8} (a_{xx} + 2 b_{xx})
\end{equation}
\begin{equation} \label{eq:cxxxy}
c_{xy} = - \frac{1}{4} c_{xx}
\end{equation}
which do not depend on the spectral tensor type $\alpha$.

Here the parameters $c_{xx}$ and $c_{xy}$, describing signature asymmetry, are zero, if the scalar part of the power spectrum  is axially symmetric too and the filter has mirror symmetry with respect to the axis $Ox$. We assume here that this is the case and we can omit the asymmetry terms.

We also need algebraic expressions for signature parameters $a_{xx}$ and $b_{xx}$. They are presented below.
\begin{equation} \label{eq:axxE}
a_{xx}^E =  \int w_{fz}(\vc{k}) \D{\vc{k}} F^2(\vc{k})
  (4\cos 2\phi \sin^2\theta)
\end{equation}
\begin{equation} \label{eq:bxxE}
b_{xx}^E =  \int w_{fz}(\vc{k}) \D{\vc{k}} F^2(\vc{k})
  (3 + \cos 2\theta - 2\cos 2\phi \sin^2\theta)
\end{equation}
\begin{equation} \label{eq:axxF}
\begin{array}{l}
a_{xx}^F = - \int w_{fz}(\vc{k}) \D{\vc{k}} F^2(\vc{k}) \\
\cdot \frac{
 4(\cos 2\phi \sin^2\theta (\cos\theta \cos\theta_\lambda + \cos\phi \sin\theta \sin\theta_\lambda)^2 + 
 \sin\theta_\lambda (\sin\theta_\lambda - \cos\phi (\cos\theta_\lambda \sin 2\theta + 2\cos\phi \sin^2\theta \sin\theta_\lambda)))
}{
 1-(\cos\theta \cos\theta_\lambda + \cos\phi \sin\theta \sin\theta_\lambda)^2
}\\
\end{array}
\end{equation}
\begin{equation} \label{eq:bxxF}
\begin{array}{l}
b_{xx}^F =  \int w_{fz}(\vc{k}) \D{\vc{k}} F^2(\vc{k}) \\
\cdot \frac{
 (2 \cos\phi \cos\theta_\lambda \sin 2\theta + (-3 + \cos 2\phi - 2\cos^2\phi \cos 2\theta) \sin\theta_\lambda)^2
}{
 4 (1-(\cos\theta \cos\theta_\lambda + \cos\phi \sin\theta \sin\theta_\lambda)^2)
} \\
\end{array}
\end{equation}



The corresponding signature parameters for the Alfv\'enic mode are as follows:
\begin{equation} \label{eq:ab_alf}
\begin{array}{l}
a_{a,xx} = a_{xx}^E-a_{xx}^F\\
b_{a,xx} = b_{xx}^E-b_{xx}^F\\
\end{array}
\end{equation}

For fast and slow modes they are as follows:
\begin{equation} \label{eq:ab_comp}
\begin{array}{l}
a_{c,xx} = a_{xx}^F\\
b_{c,xx} = b_{xx}^F\\
\end{array}
\end{equation}
\noindent with proper $F^2(\vc{k})$ (see Sect. \ref{sect:mf}).


\subsection{Impact of terms corresponding to odd powers of magnetic field} \label{sect:odd}

The direct expression for the signature $s_{xx}$ through magnetic field is as follows:
\begin{equation} \label{eq:sxx_rs}
 s_{xx} = \frac{1}{\Omega} \int_\Omega \D{\vc{R}} \left (
  \int \D{\vc{R}'} (\delta(\vc{R}'-\vc{R})-f(\vc{R}'-\vc{R})) \int \D{z'} w(z') B_x^2(\vc{r}')
 \right)^2
\end{equation}
\noindent where $\vc{R}$ is the picture-plane projection of $\vc{r}$, $z$ is the coordinate over the line of sight,  $\Omega$ is the map spot for signature calculation and $B_x=B_{0x}+B_{x,turb}$. 

As we can see here, some of the terms corresponding to odd powers of turbulent magnetic field do not diminish as we do not do ensemble averaging here.

The terms corresponding to $B_{0x}^3 B_{x,turb}(\vc{r}'')$ and $B_{0x} B_{x,turb}^3(\vc{r}'')$ are zero, because the operator $\int \D{\vc{R}'} (\delta(\vc{R}'-\vc{R})-f(\vc{R}'-\vc{R}))$ gives zero when applied to a constant\footnote{Variable $\vc{r}''$ appears when expanding square in Eq. \ref{eq:sxx_rs}}.

So, the only odd-power term giving non-zero contribution corresponds to $B_{0x} B_{x,turb}(\vc{r}') B_{x,turb}^2(\vc{r}'')$. Using the spectral representation for $B_{x,turb}$ after some algebra we can write out the following expression for its variance:
\begin{equation} \label{eq:sxx2_odd}
\begin{array}{l}
 \avg{s^2_{xx,odd}} = 16 B_{0x}^2 (2\pi)^9 
  \int \D{\vc{k}_1} F^2(\vc{k}_1) \int \D{\vc{k}_2} F^2(\vc{k}_2) \int \D{\vc{k}_3} F^2(\vc{k}_3) \cdot ( \\
  4 (j^2(|\vc{K}_2| \, R_0) +j^2(|\vc{K}_1+\vc{K}_2+\vc{K}_3| \, R_0)) 
   \ft{w}(\vc{k}_1+\vc{k}_2) \ft{w}(\vc{k}_1) \ft{w}(\vc{k}_2+\vc{k}_3) \ft{w}(\vc{k}_3) \\
  + 2 j^2(|\vc{K}_1+\vc{K}_2+\vc{K}_3| \, R_0) \ft{w}^2(\vc{k}_1+\vc{k}_2) \ft{w}^2(\vc{k}_3)
 ) \\
\end{array}
\end{equation}
\noindent where $F^2$ is the power spectrum of $B_{x,turb}$, $j(x) \equiv J_1(x)/x$, $R_0$ is a spot radius and $\ft{w}$ is a Fourier transform of
\begin{equation} \label{eq:w3}
  w(\vc{r}) \equiv (\delta(\vc{R})-f(\vc{R})) \cdot w(z)
\end{equation}
\noindent We also assume here that $w(\vc{r})$ is an even function.


Calculations using synthetic data show, that this term can be of the same order of magnitude as the quadratic one, being of unpredictable shape. This makes the investigation of quadratic term's behavior unneeded. 
We shall further assume that the linear term is dominating over these two ones. This assumption will be also checked using synthetic data.


\subsection{Fourier decomposition}

As we can see from Eqs. \ref{eq:sxx} and \ref{eq:sxy}, the linear term signatures have limited spectrum: only three first Fourier harmonics are non-zero (in addition, for the $xy$-signature the second harmonic is zero too). As practice shows, this is true for any signature, including quadratic and odd-power terms and real signatures obtained from observational data. 

So we can state the following for the signature Fourier coefficients:
\begin{equation} \label{eq:fn}
f_{^{xx}_{xy},^c_s,n} \equiv \frac{2}{\pi}\int_0^\pi {s_{^{xx}_{xy}}}(\phis)\, ^{\cos}_{\sin} (2n\phis)  \D{\phis}
\end{equation}
\begin{equation} \label{eq:fnz}
f_{*,*,n} = 0,\,n>2
\end{equation}
\begin{equation} \label{eq:fnzxy}
f_{xy,*,1} = 0
\end{equation}

In addition, the rules Eqs. \ref{eq:axxxy}, \ref{eq:bxxxy} and \ref{eq:cxxxy} can be mapped to Fourier space as follows:
\begin{equation} \label{eq:f0xxxy}
f_{xy,c,0} = \frac{1}{2}f_{xx,c,0} - f_{xx,c,2}
\end{equation}
\begin{equation} \label{eq:f2xxxy}
f_{xy,c,2} = -f_{xx,c,2}
\end{equation}
\begin{equation} \label{eq:f2xxxys}
f_{xy,s,2} = -f_{xx,s,2}
\end{equation}
Numerical calculations show, that these rules apply to quadratic and odd-power terms too. 

For signatures obtained from observational data Eqs. \ref{eq:f2xxxy} and \ref{eq:f2xxxys} hold with very high accuracy, while the disparity of Eq. \ref{eq:f0xxxy} is larger, being still small.


\section{Identification of MHD modes}

Let us introduce the parameter which is further used for MHD modes identification (see Sect. \ref{sect:sig_lin}):
\begin{equation} \label{eq:rxx}
 r_{xx} \equiv \frac{a_{xx}}{b_{xx}} = \frac{\avg{\hat{e}_{\alpha,y}^2}_\Omega}{\avg{\hat{e}_{\alpha,x}^2}_\Omega} - 1
\end{equation}

We also need its derivative over filtering scale $L_f$ of our high-pass filter for this purpose (see Eq. \ref{eq:s}). 

It is important that the line of sight window function $w(z)$, contributing to Eq. \ref{eq:wfz} is not constant, because otherwise $\tilde{w} \sim \delta(k_z)$ and dependence on $K$ in Eqs. \ref{eq:axxE}, \ref{eq:bxxE}, \ref{eq:axxF}, \ref{eq:bxxF} factorizes and cancels in Eq. \ref{eq:rxx}, what means that $r_{xx}$ does not depend on $L_f$ in this case. 

Another reason not to take such simplification is because we can reproduce observed positive $r_{xx}$ only in the case of small enough line-of-sight extent of an emitting structure with Alfv\'enic mode. 

In order to determine the classification rule we do a parameter space study within their expected parameter ranges, by numerical evaluation of Eqs. \ref{eq:ab_alf} and \ref{eq:ab_comp}.

The results of parameter space scanning are presented in Tab. \ref{tab:linsig}. The parameter ranges are: $M_a \in [0.1,0.9] $, $\theta_\lambda \in [20^\circ,80^\circ]$, $L_z \in [0.5,4]$, $L_f \in [0.25,4]$, $L=1$.
\begin{deluxetable}{lrr}[h]
\tablecaption{Fraction of negative parameters for linear term (direct calculation)\label{tab:linsig}}
\tablehead{
  \colhead{MHD mode} & \colhead{$\partial r_{xx}/ \partial L_f$} & \colhead{$r_{xx}$}
}
\startdata
Alfv\'enic         &  0.00 \% &  74.80 \% \\ 
slow               & 77.80 \% & 100.00 \% \\ 
fast, high $\beta$ & 64.00 \% & 100.00 \% \\ 
fast, low $\beta$  & 29.00 \% & 100.00 \% \\
\enddata
\end{deluxetable}

We can see, that the sign of the trend $\partial r_{xx}/ \partial L_f$ is always positive for Alfv\'enic mode and can be negative only for compressible modes.

The sign of $r_{xx}$ can be positive only for the Alfv\'enic mode, and the negative sign of $r_{xx}$ is possible for both compressible and Alfv\'enic modes.

\begin{figure}[h]
\begin{center}
\includegraphics[width=0.4\columnwidth]{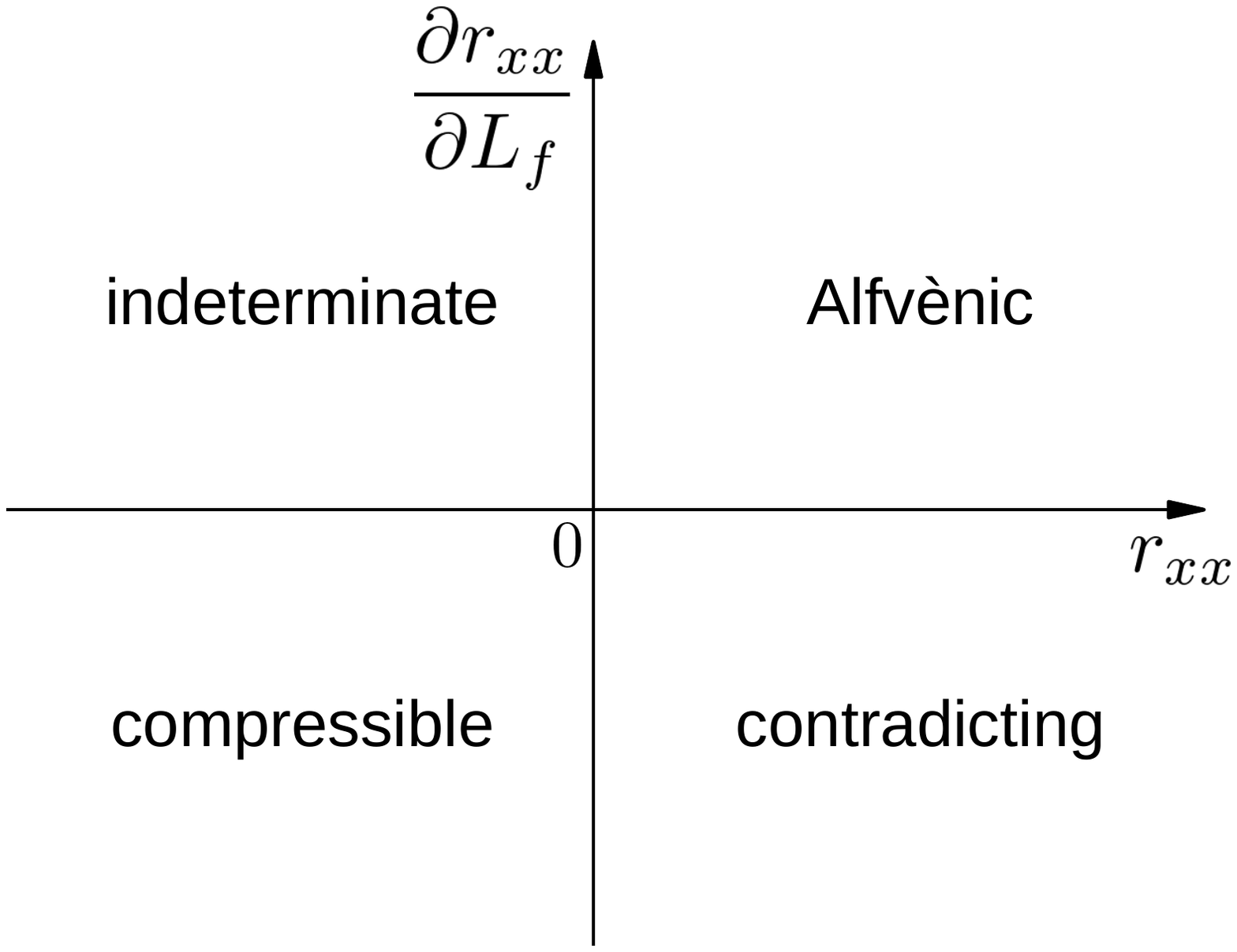}
\end{center}
\caption{MHD mode classification rule with respect to signs of parameters $\partial r_{xx}/ \partial L_f$ and $r_{xx}$. \label{fig:cls}} 
\end{figure}

This behavior allows us to identify dominating compressible and Alfv\'enic modes using parameter signs as shown on Fig. \ref{fig:cls}.

However, in reality we cannot separate the linear term and so we need to account for other signal components such as quadratic or odd-power terms, what can be done using synthetic data. 

The results of parameter space scanning for this total signal are presented in Tab. \ref{tab:totalsig}. The parameter ranges are: $M_a \in [0.1,0.7] $, $\theta_\lambda \in [20^\circ,80^\circ]$, $L_f \in [0.25,3]$, $L_z = L = 1$. The parameter space is restricted here with respect of the one corresponding to direct calculation due to high computation time.
\begin{deluxetable}{lrr}[h]
\tablecaption{Fraction of negative parameters for total signal (synthetic data)\label{tab:totalsig}}
\tablehead{
  \colhead{MHD mode} & \colhead{$\partial r_{xx}/ \partial L_f$} & \colhead{$r_{xx}$}
}
\startdata
Alfv\'enic           &  5.56 \% &  88.69 \% \\ 
slow               & 79.86 \% & 100.00 \% \\ 
fast, high $\beta$ & 62.50 \% & 100.00 \% \\ 
fast, low $\beta$  & 40.97 \% & 100.00 \% \\ 
\enddata
\end{deluxetable}

As we can see, in general our identification recipe stays valid, if we can admit a small number of false detections of compressible modes.


\section{Data processing}  \label{sect:proc}

The observed signatures and related values of $r_{xx}$ and $\Delta r_{xx}/\Delta L_f$ are calculated as follows.
We select the data inside a region with given radius and intensity level range. Then we rotate the Stokes parameters frame to the direction of mean magnetic field, what is equivalent to making the mean Q parameter maximal. After that we apply our high-pass filter to the I, Q, U maps, what is needed for improving statistics and for further classification. Then we calculate the variances of (I+Q)/2 and U/2 gradually rotating the frame of the Stokes parameters by given $\Delta \phis$. Obtained dependencies of the variances of (I+Q)/2 and U/2 on rotation angle $\phis$ form the signatures $\hat{s}_{xx}$ and $\hat{s}_{xy}$. 

If the relative variance of $\hat{s}_{xx}$ is small enough, we can guess that the signal is caused by the quadratic term of an isotropic field, so the signature is classified as "isotropic". If it is not the case, we try to figure out if the model of axially-symmetrical magnetic field statistics can be applied. 

The magnitude of the signature sine coefficients relative to constant term (asymmetry) can be used for estimating applicability of axially-symmetric magnetic field model. For the asymmetry threshold we take the relative deviation of the signal due to statistical fluctuations, which can be estimated from the magnetic field power spectrum and filter parameters, as described in Sect. \ref{sect:disp}. 

To recover the parameter $r_{xx}$ we remove the constant component equal to $\hat{s}_{xx}(90^\circ)$ from the $xx$-signature and the related value from the $xy$ one (using Eq. \ref{eq:bxxxy}) and fit the linear term theoretical signatures by the technique of least squares\footnote{Non-zero value of $\hat{s}_{xx}(90^\circ)$ can be caused by the contribution of quadratic and odd-power terms only. We also apply a threshold to this value to make sure that this impact is marginal.}. Then the procedure is repeated for different value of $L_f$ to calculate the trend $\Delta r_{xx}/\Delta L_f$ needed for the mode identification.


\subsection{Data processing validation and compressible mode observability}  \label{sect:proc_val}

\begin{figure}[h]
\begin{center}
\begin{tabular}{lll}
\includegraphics[width=0.3\textwidth]{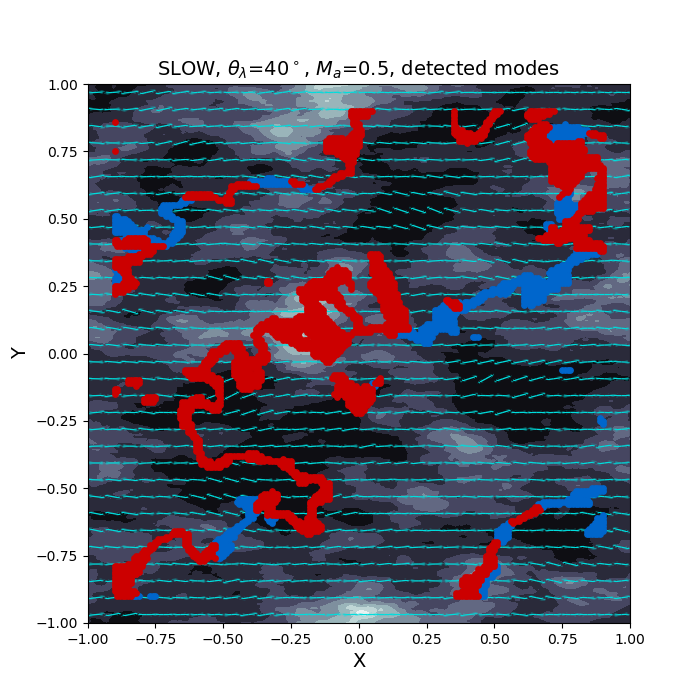} &
\includegraphics[width=0.3\textwidth]{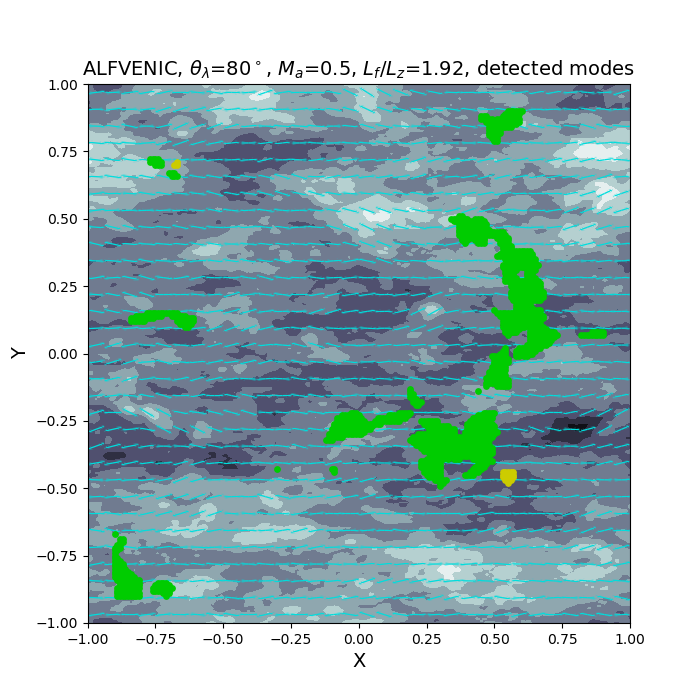} &
\includegraphics[width=0.3\textwidth]{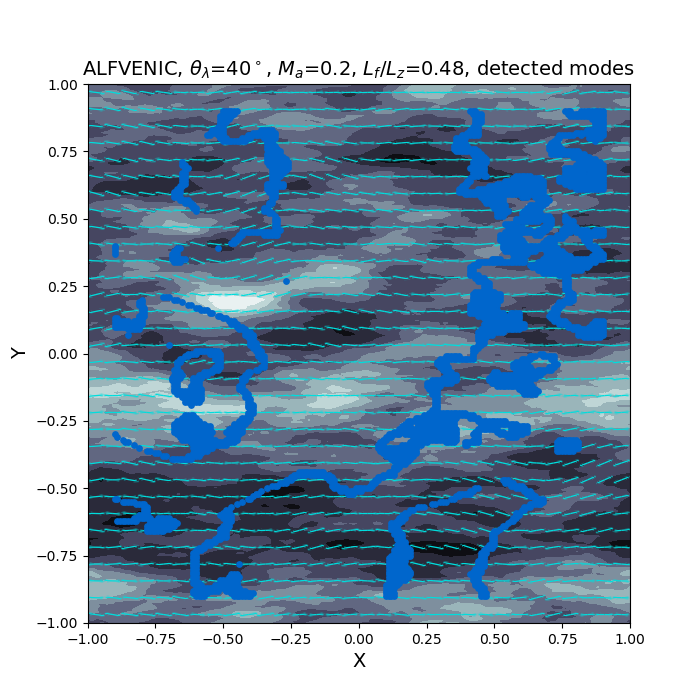} \\
\end{tabular}
\end{center}
\caption{Mode detection for Stokes parameter maps obtained from synthetic magnetic field (total signal). Left map corresponds to slow MHD mode, the compressible one.  Middle map corresponds to Alfv\'enic mode with $L_f>L_z$, $r_{xx}>0$. Right map corresponds to Alfv\'enic mode with $L_f<L_z$, $r_{xx}<0$. Red marks designate spots, classified as "compressible", green marks are "Alfv\'enic", yellow ones are "contradicting" and blue ones are "indeterminate". \label{fig:synthetic_maps}}
\end{figure}

The detection maps corresponding to synthetic magnetic field for particular MHD modes are presented on Fig. \ref{fig:synthetic_maps}. This simulation shows, that our mode identification procedure stays valid regardless of presence of contamination related to higher order terms. 

However, we must admit that we are not able to identify a MHD mode in every point of the map. This is the side effect of our identification procedure, which can be applied to spots with symmetric enough signatures only.

The identification patterns typical to these three simulated maps can be found on the real map of the North Galactic Pole region, see Fig. \ref{fig:NGP_map}. 

An important question is if we can detect compressible modes at all, taking into account that Alfv\'enic mode can dominate. 

\begin{figure}[h]
\begin{center}
\begin{tabular}{lll}
\includegraphics[width=0.3\textwidth]{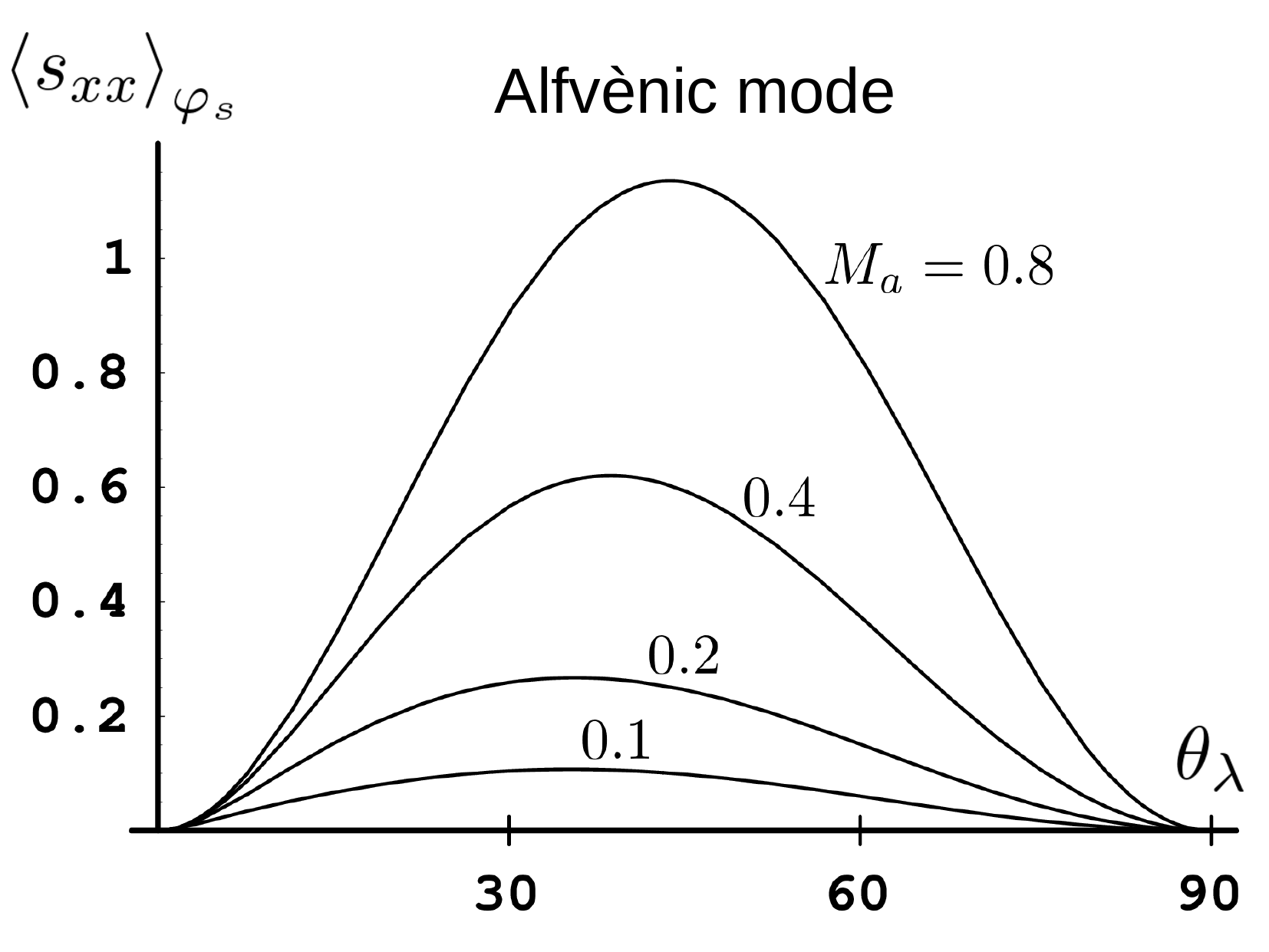} &
\includegraphics[width=0.3\textwidth]{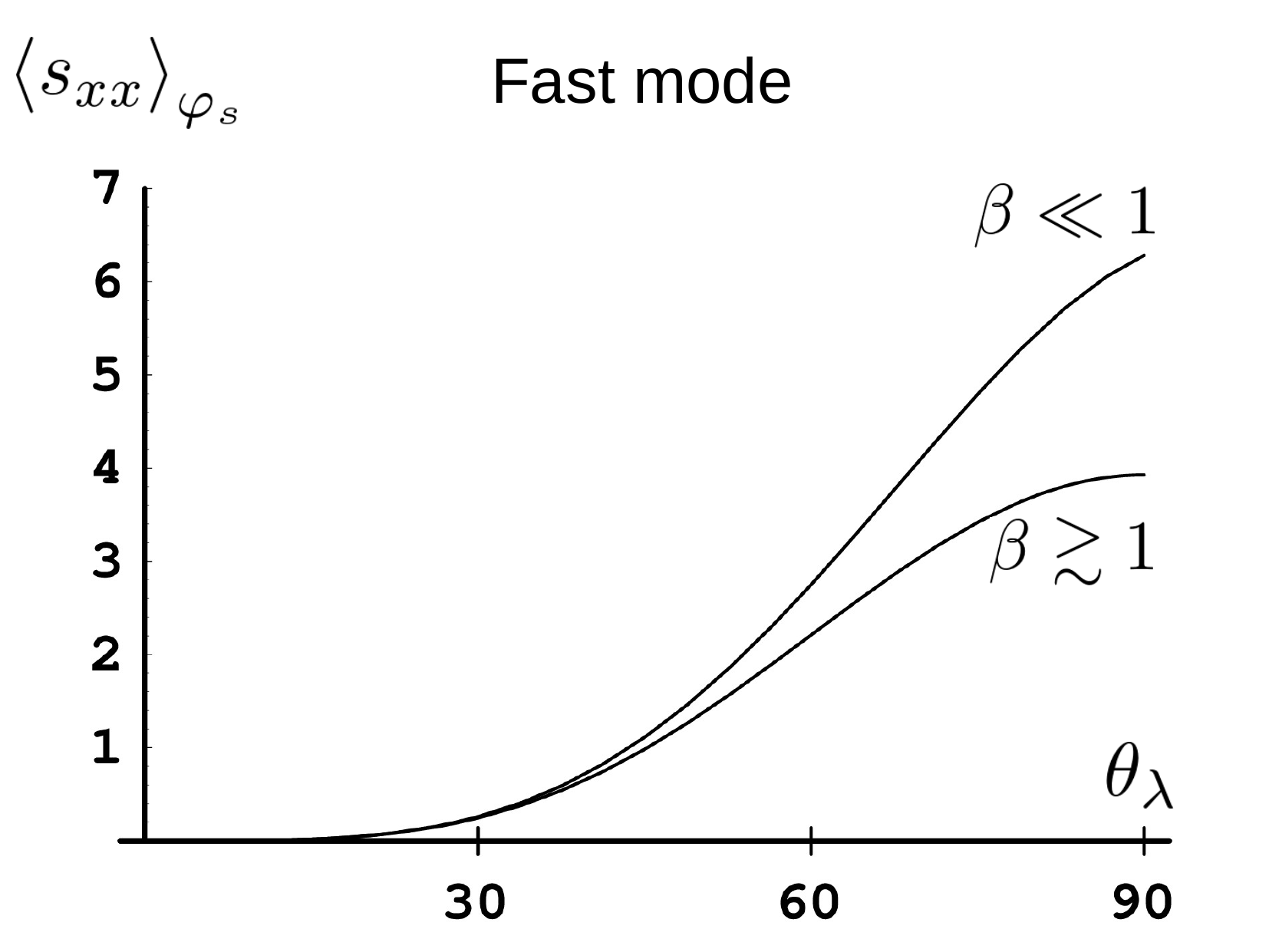} &
\includegraphics[width=0.3\textwidth]{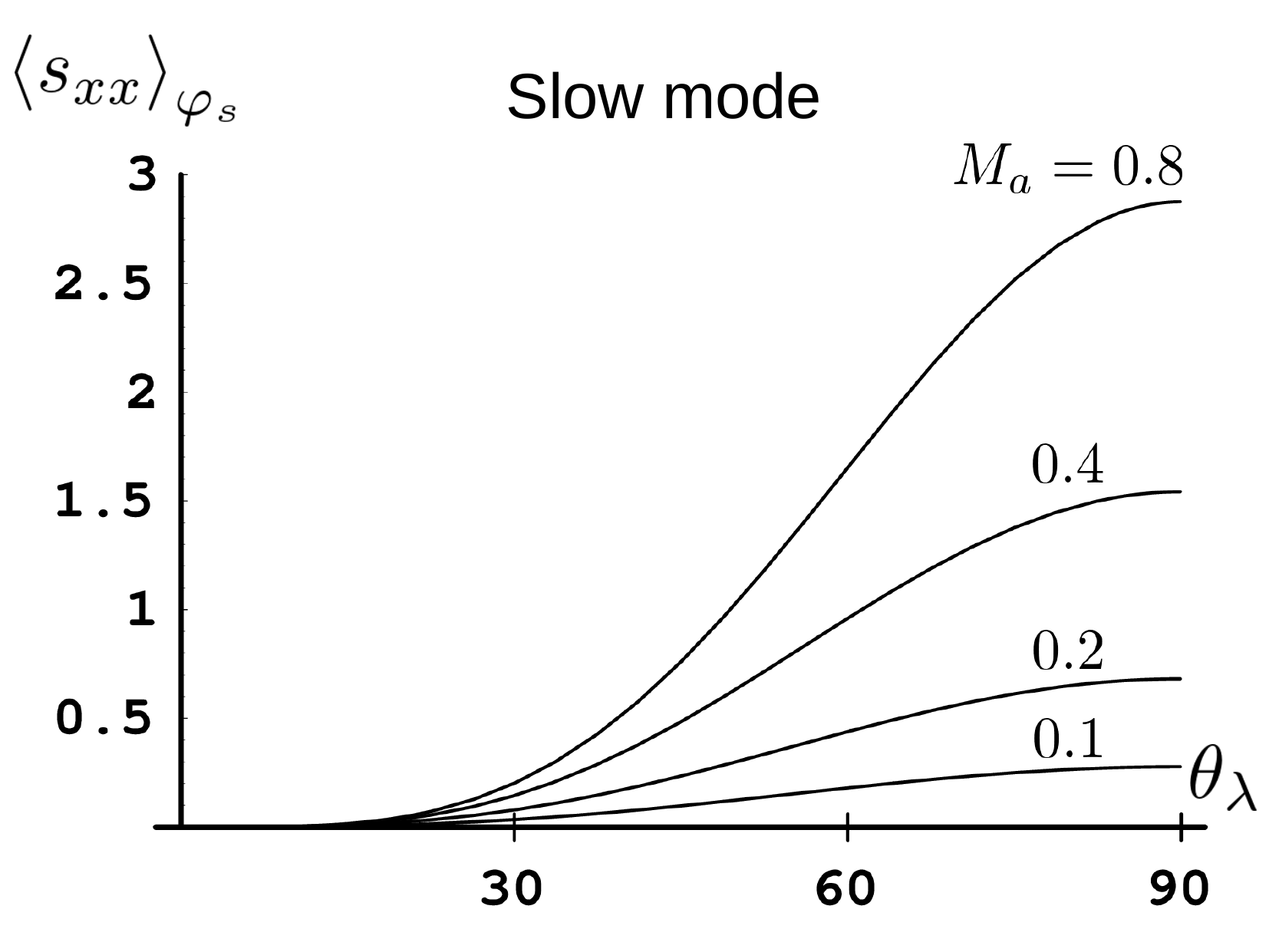} \\
\end{tabular}
\end{center}
\caption{Signal magnitude as function of $\theta_\lambda$ for different MHD modes. \label{fig:sig_behavior}} 
\end{figure}

If we plot mode signal magnitude as function of $\theta_\lambda$ for different MHD modes (see Fig. \ref{fig:sig_behavior}), we can see that at 90$^\circ$ Alfv\'enic signal is zero while slow and fast ones have their maximum. So Alfv\'enic signal can be suppressed with respect to compressible ones even if the corresponding field is stronger, what can explain observability of compressible modes. 

\section{Results of observational data processing}

\begin{figure}[h]
\begin{center}
\includegraphics[width=0.6\columnwidth]{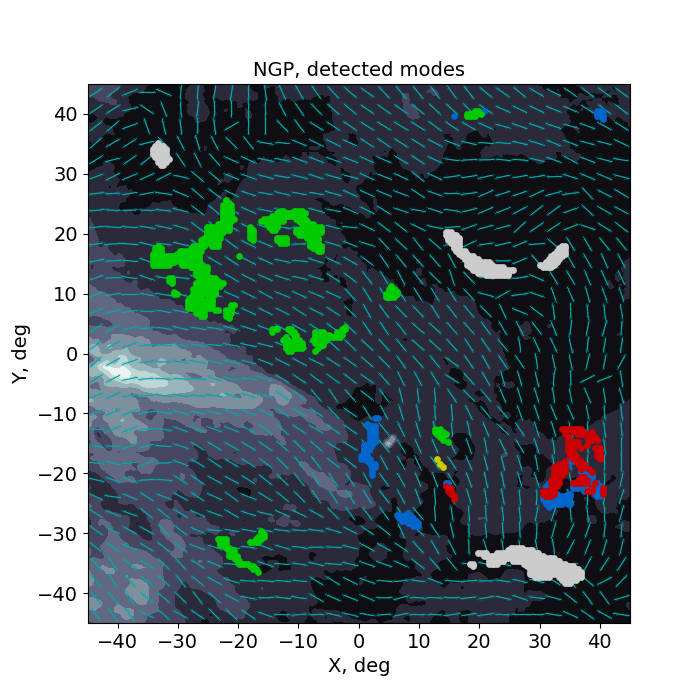}
\end{center}
\caption{Detected modes of turbulence in the vicinity of the North Galactic Pole. Green dots designate spots with domination of Alfv\'enic mode, red ones designate domination of compressible modes and white dots correspond to isotropic turbulence. Blue dots correspond to spots with indeterminate status, yellow dots represent data contradicting our analysis. Spot radius is 9$^\circ$.5. \label{fig:NGP_map}} 
\end{figure}

\begin{figure}[h]
\begin{center}
\includegraphics[width=0.3\columnwidth]{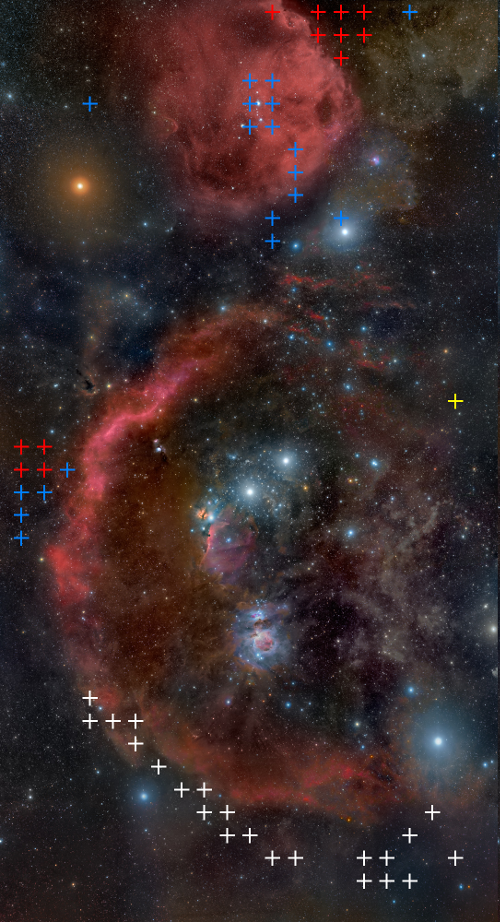}
\end{center}
\caption{Detected modes of turbulence in the Orion molecular cloud complex. Red crosses designate domination of the compressible modes and white crosses correspond to the isotropic turbulence. Blue crosses correspond to spots with indeterminate status, yellow crosses represent contradicting data. Spot radius is 6$^\circ$.4 or 50 pc, map size is $14^\circ \times 24^\circ$. \label{fig:Orion_map}} 
\end{figure}

\begin{figure}[h]
\begin{center}
\includegraphics[width=0.5\columnwidth]{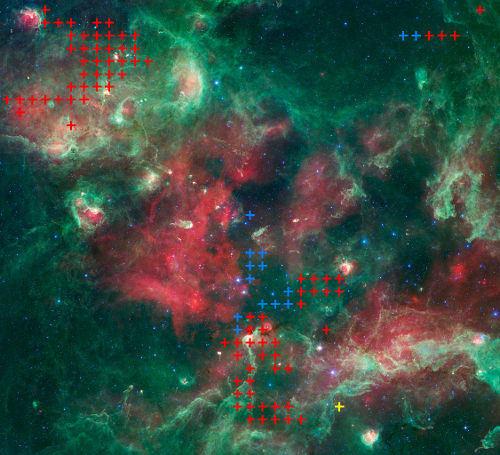}
\end{center}
\caption{Detected modes of turbulence in Cygnus X star formation region. Red crosses designate domination of the compressible modes, blue crosses correspond to spots with indeterminate status, yellow crosses represent contradicting data. Spot radius is 1$^\circ$.05 or 26 pc, map size is $5^\circ \times 5^\circ$. \label{fig:Cygnus_map}} 
\end{figure}

In this work we investigate the presence of compressible MHD modes in the turbulent interstellar medium using polarization maps of radio-frequency synchrotron emission. We study three Galactic regions of the ISM: the vicinity of the North Galactic Pole (NGP), the Orion molecular cloud complex and the star-forming complex Cygnus X. We use the DRAO 1.4 GHz polarization survey with angular resolution 36$'$ \citep{W06} for the NGP region and Orion complex, and employ Urumqi 6cm polarization survey with resolution 9$'$ \citep{X11} for Cygnus X complex, see Figures \ref{fig:NGP_map}, \ref{fig:Orion_map} and \ref{fig:Cygnus_map}.

All three regions display presence of compressible MHD modes. The presence of "contradicting" data is quite small, what also confirms validity of our technique.

Faraday rotation for NGP map spots with detected modes is $10^\circ \div 15^\circ$, and regarding Sect. \ref{sect:faraday} has marginal influence\footnote{This Faraday rotation is estimated from all-sky Faraday rotation map by \cite{Opp15}.}.

The dominance of compressible turbulence in star-forming regions is consistent with the picture of turbulence injection by the highly supersonic flows (inside the dense and cold molecular clouds), generated by supernova explosions and outflows of YSOs.

Another option for observing a compressible signal could be suppressing the Alfv\'enic signal with respect to compressible ones due to a small angle between mean magnetic field and picture plane, as shown in Sect. \ref{sect:proc_val}. This could be the case for detection of a compressible mode in the NGP region.

\newpage

\section{Assumptions summary}

We use the following assumptions for our technique.

\begin{itemize}
\item 
Lines of sight within the observed structure can be considered parallel. For the objects Orion and Cygnus X it is true. For the NGP region this approximation still can be used, because only anisotropy effects are relevant for us, while convergence of lines of sight is responsible for change of scales only.
\item
Faraday rotation influence can be considered marginal. For the case of NGP region the observed values of Faraday rotation do not affect our result, as estimated in Sect. \ref{sect:faraday}. For Orion and Cygnus X objects Faraday rotation within the object is unknown.
\item
The relativistic electron spectral index $\alpha = 3$ is assumed in calculation of Stokes parameters. As shown in \cite{LP12}, the change of the index does not change the spatial properties of correlations. It is also numerically checked in Sect. \ref{sect:alpha}, showing that deviation of $\alpha$ from 3 does not affect our results.
\item
Scanning contamination does not change our results. As shown in Sect. \ref{sect:sca}, it does not increase the number of false detections.
\item
Axially symmetric magnetic field model can be applicable. This axially symmetric model of MHD turbulence is suggested in \cite{LP12} on the basis of earlier theoretical and numerical work. The limitations of the model come from the variations of the magnetic field direction along the line of sight. In this work we check for applicability of this model by checking the symmetry of obtained signatures. As shown in Sect. \ref{sect:proc_val} this data selection does not affect the validity of our results. The influence of symmetry axis variation on $r_{xx}$ is also discussed in Sect. \ref{sect:fluct}.
\item
Results obtained for linear term are applicable for total signal. This assumption is proved numerically in Sect. \ref{sect:proc_val} and Sect. \ref{sect:ver}, with exception of a small number ($\sim$6\%) of compressible mode false detections due to statistical effects.
\item
Compressible signal can dominate. Alfv\'enic linear term signal can be suppressed with respect to compressible ones for $\theta_\lambda \sim 90^\circ$, as shown in Sect. \ref{sect:proc_val}. However the Alfv\'enic quadratic term is not suppressed in this case, but it has much lower mean magnitude and hardly can suppress compressible signal. 
\end{itemize}

\section{Relation to previous publications}

The incomplete version of this technique was published by \cite{Zh20}, employing the mode identification rule based on the value of $r_{xx}$ only. This rule results in 34\% of false detections of compressible modes while scanning the parameter space with numerical evaluation of analytical expression for $r_{xx}$. The same test for the technique presented here shows no false detections. 

\section{The code}

Parameter space exploration code with numerical evaluation of analytical expression for $r_{xx}$ is provided along with this paper. Data processing and synthetic field generation code is available by request (chepurnov@arcor.de).

\section{Acknowledgments}

Authors are particularly grateful to Alex Lazarian for useful discussions and to Huirong Yan for partial financial support of this research.

\newpage

\appendix


\section{Verification with synthetic data} \label{sect:ver}

We have performed the comparison of directly calculated $r_{xx}$ using Eqs. \ref{eq:ab_alf}, \ref{eq:ab_comp} and \ref{eq:rxx}, with the ones recovered from the synthetic data. 

This way we can check both our analytic results for $r_{xx}$ and the data processing procedure. The comparison shows good correspondence including the case when non-linear terms have been taken into account, see Tab. \ref{tab:comp}.  

\begin{deluxetable}{cccc}[h]
\tablecaption{Comparison of the directly calculated parameter $r_{xx}$, with the one recovered from synthetic data for different MHD modes \label{tab:comp}}
\tablehead{
	\colhead{MHD mode} & \colhead{$r_{xx}$, direct calculation} & \colhead{$r_{xx}$, synthetic} & \colhead{$r_{xx}$, synthetic} \\
	\colhead{\,}       & \colhead{linear term}                  & \colhead{linear term}         & \colhead{total signal} \\
}
\startdata
Alfv\'enic         & -0.258 & -0.240 & -0.274 \\
slow               & -0.959 & -0.958 & -0.958 \\
fast, low $\beta$  & -0.429 & -0.419 & -0.415 \\
fast, high $\beta$ & -0.707 & -0.701 & -0.698 \\
\enddata
\tablecomments{The model parameters are as follows: $M_a=0.25$, $\theta_\lambda=70^\circ$, $L=L_z=1$, $L_f=2$, $L_b=0.2$ (the latter is the beam scale).}
\end{deluxetable}


\section{The impact of the scanning contamination} \label{sect:sca}

Most of present polarization data suffer from contamination in intensity channel, which is caused by errors when assembling the map from individual scans. Practice shows, that polarization angle is more reliable parameter than intensity in this case. 

If we assume that polarization angle is not contaminated, we can adopt the model when the observed Stokes parameter is a product of an uncontaminated one and the common positive contamination factor $\alpha$:
\begin{equation}
\begin{array}{l}
\hat{I} = \alpha I \\
\hat{U} = \alpha U \\
\hat{Q} = \alpha Q \\
\end{array}
\end{equation}
Then for the signal $S\equiv(I+Q)/2$ we have the same relation:
\begin{equation}
\hat{S} = \alpha S
\end{equation}
Let us write out the correspondent variance:
\begin{equation}
D \hat{S} = \int \D{\vc{K}} (\ft{\alpha}^2 \star F^2_s) \ft{f}^2 = \int \D{\vc{K}}  F^2_s (\ft{\alpha}^2 \star \ft{f}^2) 
\end{equation}
\noindent where tilde denotes Fourier transform, $f$ is our high-pass filter and $F^2_s$ is the signal emissivity power spectrum. Therefore the expression
\begin{equation}
\ft{f}_\alpha^2 = \ft{\alpha}^2 \star \ft{f}^2
\end{equation}
\noindent gives us the modified filter. Let us model the scanning contamination factor as follows:
\begin{equation}
\alpha = 1+\sqrt{2}\delta\alpha\cos \vc{K}_\Delta \vc{R} 
\end{equation}
Then, the correspondent squared Fourier transform is as follows:
\begin{equation}
\ft{\alpha}^2 \sim \delta(\vc{K}) + \frac{\delta\alpha}{2}(\delta(\vc{K}-\vc{K}_\Delta)+\delta(\vc{K}+\vc{K}_\Delta))
\end{equation}
\noindent which gives us the variance:
\begin{equation}
D \hat{S} \sim \int \D{\vc{K}} F^2_s 
\cdot \ft{f}^2 + \frac{\delta\alpha}{2}\int \D{\vc{K}} F_s^2 \cdot (\ft{f}^2(\vc{K}-\vc{K}_\Delta)+\ft{f}^2 (\vc{K}+\vc{K}_\Delta)) 
\end{equation}
\noindent where the first term gives us the unchanged signature and the second one is the signature part affected by contamination. 

While the filter term in the contaminated signature is non-zero at low frequencies, the contaminated signature is more affected by the statistical noise, what makes the calculated $r_{xx}$ less reliable. 

Another contaminating factor is the constant component, modulated by scanning. It produces the false signal with $r_{xx} = -1$ (this value does not depend on any parameters). 

To estimate impact of these factors we have performed calculations using synthetic data.

The results of exploring of parameter space are presented in Tables \ref{tab:contdrxx} and \ref{tab:contrxx}. The parameter ranges are: $M_a \in [0.1,0.7] $, $\theta_\lambda \in [20^\circ,80^\circ]$, $L_f \in [0.25,3]$, $L_z = L = 1$.\begin{deluxetable}{lrrrr}[h]
\tablecaption{Fraction of negative $\partial r_{xx}/ \partial L_f$ for contaminated total signal \label{tab:contdrxx}}
\tablehead{
  \colhead{MHD mode} & \colhead{cont. level 0 \%} & \colhead{cont. level 5 \%} & \colhead{cont. level 10 \%} & \colhead{cont. level 20 \%}
}
\startdata
Alfv\'enic          &  5.56 \% &  4.17 \% &  3.47 \% & 2.78 \% \\ 
slow                & 79.86 \% & 31.94 \% & 10.42 \% & 0.00 \% \\ 
fast, high $\beta$  & 62.50 \% & 25.69 \% &  6.94 \% & 0.00 \% \\ 
fast, low $\beta$   & 40.97 \% & 15.97 \% &  4.17 \% & 0.00 \% \\ 
\enddata
\end{deluxetable}
\begin{deluxetable}{lrrrr}[h]
\tablecaption{Fraction of negative $r_{xx}$ for contaminated total signal \label{tab:contrxx}}
\tablehead{
  \colhead{MHD mode} & \colhead{cont. level 0 \%} & \colhead{cont. level 5 \%} & \colhead{cont. level 10 \%} & \colhead{cont. level 20 \%}
}
\startdata
Alfv\'enic          &  88.69 \% &  96.43 \% &  99.40 \% & 100.00 \% \\ 
slow                & 100.00 \% & 100.00 \% & 100.00 \% & 100.00 \% \\ 
fast, high $\beta$  & 100.00 \% & 100.00 \% & 100.00 \% & 100.00 \% \\ 
fast, low $\beta$   & 100.00 \% & 100.00 \% & 100.00 \% & 100.00 \% \\ 
\enddata
\end{deluxetable}

As we can see here, scanning contamination does not increase the number of compressible mode false detections. However it decreases the chances to identify compressible and Alfv\'enic modes.

\section{Stokes parameters for different slope of electron energy spectrum} \label{sect:alpha}

Expressions for Stokes parameters Eqs. \ref{eq:i} - \ref{eq:u} are written for the case of relativistic electron spectral index $\alpha$ equal to 3. Let us consider the different case. 

For arbitrary $\alpha$ the emissivities of Stokes parameters are as follows:
\begin{equation}
\begin{array}{l}
\eps_{0,Q} = |B_\bot|^\gamma \\
\eps_{0,U} = 0 \\
\end{array}
\end{equation}
\noindent where $\gamma=(\alpha+1)/2$ and the transverse magnetic field component is aligned over x-axis of the Stokes parameter frame (see \cite{Pan16}).

Let us rotate the Stokes frame by angle $\vphi$:
\begin{equation}
\begin{array}{l}
\eps_Q = |B_\bot|^\gamma \cos 2\vphi = B_\bot^2 (\cos^2 \vphi - \sin^2 \vphi) \cdot |B_\bot|^{\gamma-2} \\
\eps_U = |B_\bot|^\gamma \sin 2\vphi = 2 B_\bot^2 \cos \vphi \sin \vphi \cdot |B_\bot|^{\gamma-2} \\
\end{array}
\end{equation}

And, finally:
\begin{equation} \label{eq:gamma}
\begin{array}{l}
\eps_Q = (B_x^2-B_y^2) \cdot |B_\bot|^{\gamma-2} \\
\eps_U = 2 B_x B_y \cdot |B_\bot|^{\gamma-2} \\
\end{array}
\end{equation}
\noindent where $(x,y)$ is the rotated coordinate system.

\begin{figure*}[h]
\begin{center}
\begin{tabular}{ll}
\includegraphics[width=0.4\textwidth]{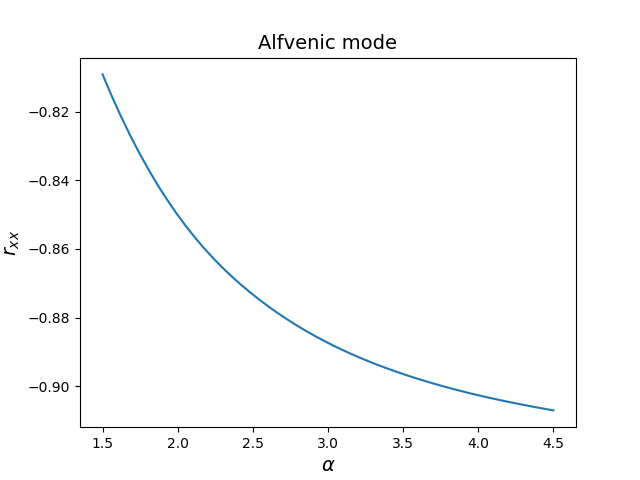} &
\includegraphics[width=0.4\textwidth]{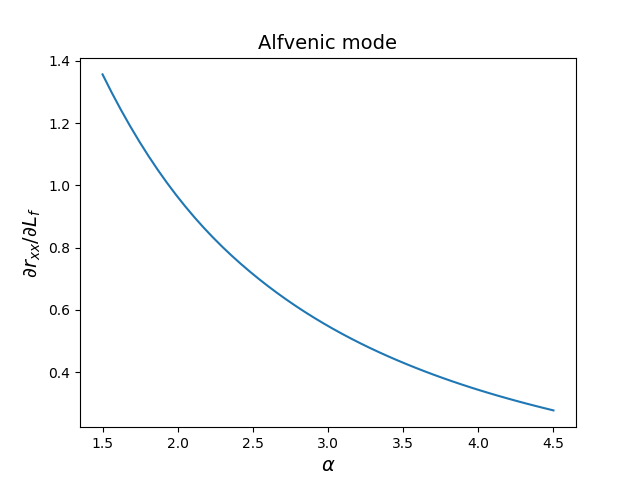} \\
\includegraphics[width=0.4\textwidth]{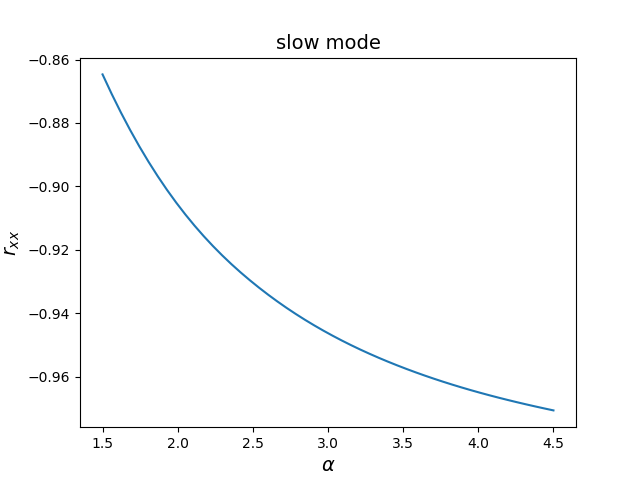} &
\includegraphics[width=0.4\textwidth]{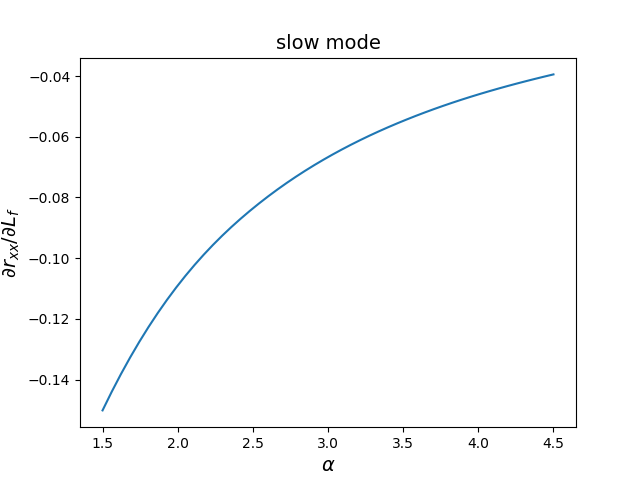} \\
\end{tabular}
\end{center}
\caption{Example dependency of parameters $r_{xx}$ (left) and $\partial r_{xx}/ \partial L_f$ (right) for Alfv\'enic (top) and slow (bottom) modes on relativistic electron spectral index $\alpha$. Other calculation parameters are $M_a=0.2$, $\theta_\lambda=40^\circ$, $L_f/L_z=0.36$.\label{fig:alpha}} 
\end{figure*}

For the estimation of $\alpha$-dependence of the parameters $r_{xx}$ and $\partial r_{xx}/ \partial L_f$ we calculated Stokes parameters from synthetic magnetic field using Eq. \ref{eq:gamma}. These simulations show, that for $\alpha \in [1.5,4.5]$ parameters $r_{xx}$ and $\partial r_{xx}/ \partial L_f$ do not change their sign, so our identification recipe stays valid in this range, see Fig. \ref{fig:alpha}

It was shown in LP12 that the correlations of synchrotron emission for an arbitrary index of cosmic rays can be presented as a combination of the prefactor that is a function of $\gamma$ and the part that depends on $R$ that is calculated for $\gamma=2$. Our results in these figures can be expressed through this prefactor obtained in LP12.

\section{Estimation of the Faraday rotation impact} \label{sect:faraday}

For the estimation of the Faraday rotation impact we calculated Stokes parameters (Eqs. \ref{eq:i}, \ref{eq:q} and \ref{eq:u}) from synthetic magnetic field, applying Faraday rotation numerically. Then the obtained maps were used to calculate the related parameters $r_{xx}$ and $\partial r_{xx}/ \partial L_f$.

\begin{figure*}[h]
\begin{center}
\begin{tabular}{ll}
\includegraphics[width=0.4\textwidth]{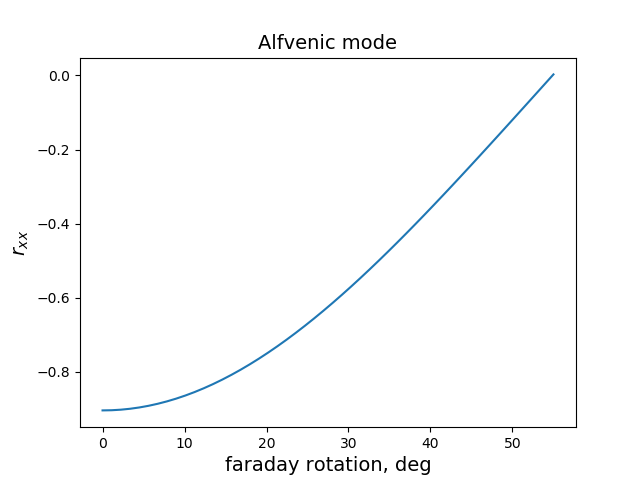} &
\includegraphics[width=0.4\textwidth]{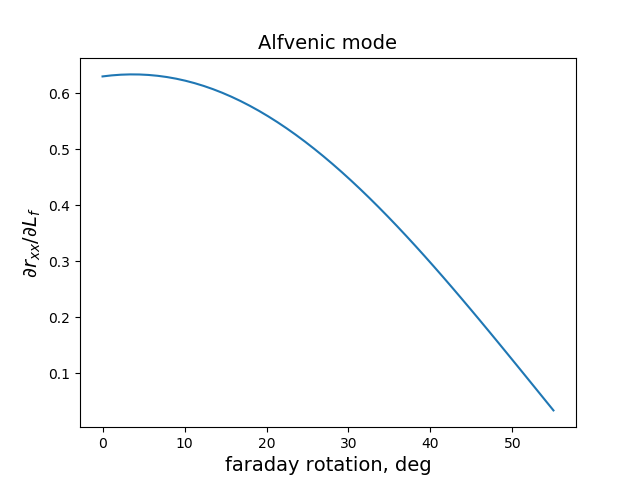} \\
\includegraphics[width=0.4\textwidth]{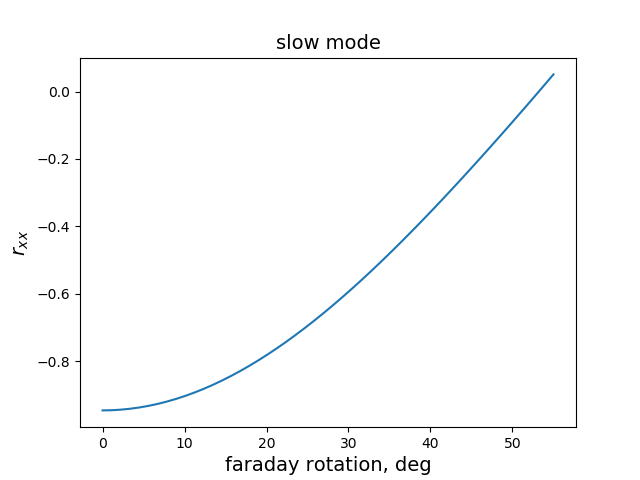} &
\includegraphics[width=0.4\textwidth]{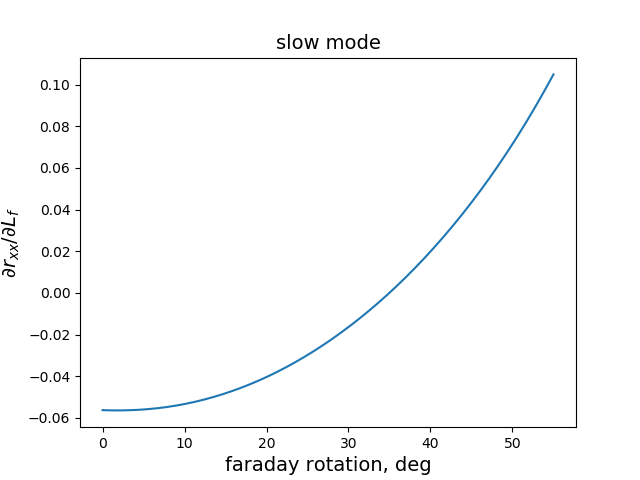} \\
\end{tabular}
\end{center}
\caption{Example of Faraday rotation impact on parameters $r_{xx}$ (left) and $\partial r_{xx}/ \partial L_f$ (right) for Alfv\'enic (top) and slow (bottom) modes. Other calculation parameters are $M_a=0.2$, $\theta_\lambda=40^\circ$, $L_f/L_z=0.36$.\label{fig:faraday}} 
\end{figure*}

These simulations show, that for Faraday rotation up to  $\sim 30^\circ$ these parameters do not change their sign, so our identification recipe stays valid in this range, see Fig. \ref{fig:faraday}.

\vspace{1cm}

\section{Accounting for fluctuations of the symmetry axis} \label{sect:fluct}

Let us model the influence of the symmetry axis fluctuations on the signature linear term as follows:
\begin{equation} \label{eq:sxx_fluct}
s^{fluct}_{xx}(\phis)= \frac{1}{2 \Delta \phis} \int_{-\Delta \phis}^{\Delta \phis} s_{xx}(\phis-\phis') \D{\phis'}
\end{equation}
\noindent where $\Delta \phis$ is the fluctuation angle. 

After such operation the signature value at $90^\circ$ is not zero any more:
\begin{equation} \label{eq:sxx_fluct_90}
s^{fluct}_{xx}(90^\circ)= \frac{1}{32\Delta \phis} (4(a_{xx} + 4b_{xx})\Delta \phis - 8b_{xx}\sin 2\Delta \phis - a_{xx}\sin 4\Delta \phis)
\end{equation}

If we subtract this value from Eq. (\ref{eq:sxx_fluct}), we have the signature of the form Eq. (\ref{eq:sxx}) with the parameters expressed through original $a_{xx}$ and $b_{xx}$:
\begin{equation} \label{eq:abhat}
\begin{array}{l}
\hat{a}_{xx} = a_{xx} \frac{\sin 4\Delta \phis}{4\Delta \phis} \\ 
\hat{b}_{xx} = b_{xx} \frac{\sin 2\Delta \phis}{2\Delta \phis} 
\end{array}
\end{equation}

And, finally:
\begin{equation} \label{eq:rxx_obs}
\hat{r}_{xx} = r_{xx} \frac{\sin 4\Delta \phis}{2 \sin 2\Delta \phis} 
\end{equation}

I.e. the observed $\hat{r}_{xx}$ differs from the original $r_{xx}$ by a factor which does not change its sign if $|\Delta \phis|<45^\circ$, so within this range our identification recipe stays valid.  


\section{Statistical error of the dispersion of a scalar random field} \label{sect:disp}

Let us take a homogeneous random field, which admits a spectral representation  

\begin{equation}
s(\vc{R}) = \int e^{i\vc{K}\vc{R}} F(\vc{K}) \xi(\vc{K}) \sqrt{\D{\vc{K}}}.
\end{equation}

For its dispersion $D_\Omega$, estimated over a limited area $\Omega$, after some algebra we can write the following expression:

\begin{equation} \label{eq:d}
D_\Omega = \int F^2(\vc{K})(1-\Pi^2_\Omega(\vc{K})) \D{\vc{K}}
\end{equation}

\noindent where

\begin{equation}
\Pi_\Omega(\vc{K}) = \frac{1}{\Omega} \int_\Omega e^{i\vc{K}\vc{R}} \D{\vc{R}}
\end{equation}

Then the variance of $D_\Omega$ can be evaluated as follows:

\begin{equation} \label{eq:dd}
\sigma^2_D = 2 \int F^2(\vc{K}) \D{\vc{K}} \int F^2(\vc{K}') \D{\vc{K}'} \cdotp (\Pi_\Omega(\vc{K}+\vc{K}') - \Pi_\Omega(\vc{K})\Pi_\Omega(\vc{K}'))^2
\end{equation}


\section{Spectral representation of a random field} \label{sect:sr}

The following considerations, formulated in \cite{Ch21}, are based on results obtained in \cite{IR70}, \cite{R90} and \cite{Ch98}.


\subsection{Scalar field} \label{sect:sf}

Let us write the correlation function of a homogeneous random field $\rho(\vc{r})$ through its power spectrum:
\begin{equation} \label{eq:scf}
C(\vc{r}) \equiv \avg{\rho^*(\vc{0})\rho(\vc{r})} = \int e^{i\vc{k}\vc{r}} \mcl{F}(d\vc{k}) = \int e^{i\vc{k}\vc{r}} F^2(\vc{k}) \D{\vc{k}} 
\end{equation}
\noindent where $\mcl{F}(\cdot)$ is a correlation spectral measure and $F^2$ is the correspondent power spectrum.

Every homogeneous field admits a spectral representation as follows:
\begin{equation} \label{eq:ssr_orig}
\rho(\vc{r}) = \int e^{i\vc{k}\vc{r}} \Phi(d\vc{k}) 
\end{equation}
\noindent where $\Phi(\cdotp)$ is a complex random measure in $\mathbb{R}^3$, satisfying\footnote{It follows from Eq. (\ref{eq:sphi}) that this field is Gaussian.}
\begin{equation} \label{eq:sphi}
\avg{\Phi(A)\Phi^*(B)} = \mcl{F}(A\cap B) 
\end{equation}

Consequently, measure elements must conform the following symbolic rule:
\begin{equation} \label{eq:sphi_symb_orig}
\avg{\Phi(d\vc{k})\Phi^*(d\vc{k}')} = \delta_{\vc{k}\vc{k}'} \mcl{F}(d\vc{k}) =  \delta_{\vc{k}\vc{k}'} F^2(\vc{k}) \D{\vc{k}}
\end{equation}

We would like to improve the notation Eq. (\ref{eq:ssr_orig}). It would be interesting to expose the internal structure of the spectral measure element. If we introduce the complex random field $\xi$ as follows
\begin{equation} \label{eq:sphi_symb}
\Phi(d\vc{k}) = F(\vc{k}) \, \xi(\vc{k}) \sqrtd{k}
\end{equation}
\begin{equation} \label{eq:xi_corr}
\avg{\xi(\vc{k})\xi^*(\vc{k}')} = \delta_{\vc{k}\vc{k}'},
\end{equation}
\noindent we satisfy Eq. (\ref{eq:sphi_symb_orig}). Then, Eq. (\ref{eq:ssr_orig}) can be rewritten in the following form:
\begin{equation} \label{eq:ssr}
\rho(\vc{r}) = \int e^{i\vc{k}\vc{r}} F(\vc{k}) \, \xi(\vc{k}) \sqrtd{k} 
\end{equation}
\noindent where $\xi(\vc{k})$ conforms Eq. (\ref{eq:xi_corr}) and $F(\vc{k})$ is square root of the power spectrum. In addition, if $\rho \in \mathbb{R}$,
\begin{equation} \label{eq:sxi}
\xi(-\vc{k}) = \xi^*(\vc{k}).
\end{equation}


\subsection{Vector field}

Let us generalize Eq. (\ref{eq:ssr}) for the case of a vector field.

Let us write the correlation function of a homogeneous vector random field $u_i(\vc{r})$ through its power spectrum:
\begin{equation} \label{eq:vcf}
C_{ij}(\vc{r}) \equiv \avg{u_i^*(\vc{0})u_j(\vc{r})} = \int e^{i\vc{k}\vc{r}} \mcl{F}_{ij}(d\vc{k}) = \int e^{i\vc{k}\vc{r}} F^2(\vc{k}) T_{ij}(\vch{k}) \D{\vc{k}} 
\end{equation}
\noindent where $\mcl{F}_{ij}(\cdot)$ is a tensor correlation spectral measure, $F^2$ is the scalar part of a power spectrum and $T_{ij}$ is a spectral tensor, satisfying the equations
\begin{equation} \label{eq:vt}
T_{ij}(\vch{k}) = T_{il}(\vch{k})T_{lj}(\vch{k})
\end{equation}
\begin{equation} \label{eq:tsym}
T_{ij}(\vch{k}) = T_{ji}(\vch{k})
\end{equation}

Eqs. (\ref{eq:vt}) and (\ref{eq:tsym}) indicate that $T_{ij}$ is an orthogonal projector, see for example \cite{TB97}.

The spectral representation of the field itself can be written as follows:
\begin{equation} \label{eq:vsr_orig}
u_i(\vc{r}) = \int e^{i\vc{k}\vc{r}} \Phi_i(d\vc{k}) 
\end{equation}
\noindent where $\Phi_i(\cdotp)$ is a complex vector random measure in $\mathbb{R}^3$, satisfying
\begin{equation} \label{eq:vphi}
\avg{\Phi_i(A)\Phi_j^*(B)} = \mcl{F}_{ij}(A\cap B) 
\end{equation}

The correspondent symbolic rule for $\Phi_i(d\vc{k})$ can be written in the following form:
\begin{equation} \label{eq:vphi_symb_orig}
\avg{\Phi_i(d\vc{k})\Phi_j^*(d\vc{k}')} = \delta_{\vc{k}\vc{k}'} \mcl{F}_{ij}(d\vc{k}) =  \delta_{\vc{k}\vc{k}'} F^2(\vc{k}) T_{ij}(\vch{k}) \D{\vc{k}}
\end{equation}

Then the explicit form of the spectral measure element can be represented as follows:
\begin{equation} \label{eq:vphi_symb}
\Phi_i(d\vc{k}) = F(\vc{k}) \, T_{ij}(\vch{k}) \xi_j(\vc{k}) \sqrtd{k},
\end{equation}
\noindent where the complex random field $\xi_i(\vc{k})$ must conform:
\begin{equation} \label{eq:vxicorr}
\avg{\xi_i(\vc{k})\xi_j^*(\vc{k}')} = \delta_{ij} \delta_{\vc{k}\vc{k}'}.
\end{equation}
\noindent This satisfies Eq. (\ref{eq:vphi_symb_orig}). Finally, Eq. (\ref{eq:vsr_orig}) can be rewritten as follows:
\begin{equation} \label{eq:vsr}
u_i(\vc{r}) = \int e^{i\vc{k}\vc{r}} F(\vc{k}) \,  T_{ij}(\vch{k}) \xi_j(\vc{k}) \sqrtd{k} 
\end{equation}
\noindent where $\xi_i(\vc{k})$ conforms Eq. (\ref{eq:vxicorr}), $F(\vc{k})$ is square root of the scalar part of the power spectrum and the spectral tensor $T_{ij}(\vch{k})$ conforms Eqs. (\ref{eq:vt}) and (\ref{eq:tsym}). 

If the field is real, we also have
\begin{equation} \label{eq:vxisym}
\xi_i(-\vc{k}) = \xi_i^*(\vc{k}).
\end{equation}


\subsection{Geometrical properties of spectral tensors} \label{sect:geom}

Let us consider the following spectral tensors for isotropic potential field and axially symmetrical compressible and Alfv\'enic fields described in \cite{LP12}:
\begin{equation} \label{eq:Tp}
T_{p,ij}(\vch{k}) = \hat{k}_i\hat{k}_j
\end{equation}
\begin{equation} \label{eq:Tc}
T_{c,ij}(\vch{k}) = \frac{
(\vch{k}\vch{\lambda})^2\hat{k}_i\hat{k}_j+\hat{\lambda}_i\hat{\lambda}_j-(\vch{k}\vch{\lambda})(\hat{k}_i\hat{\lambda}_j+\hat{\lambda}_i\hat{k}_j)
}{1-(\vch{k}\vch{\lambda})^2}
\end{equation}
\begin{equation} \label{eq:Ta}
T_{a,ij}(\vch{k}) = \delta_{ij}-\hat{k}_i\hat{k}_j - \frac{
(\vch{k}\vch{\lambda})^2\hat{k}_i\hat{k}_j+\hat{\lambda}_i\hat{\lambda}_j-(\vch{k}\vch{\lambda})(\hat{k}_i\hat{\lambda}_j+\hat{\lambda}_i\hat{k}_j)
}{1-(\vch{k}\vch{\lambda})^2}
\end{equation}
\noindent where $\vch{\lambda}$ defines the symmetry axis.

As these tensors satisfy Eqs. (\ref{eq:vt}) and (\ref{eq:tsym}), they are orthogonal projectors. 

Their ranges are one-dimensional and their range basis vectors form in $\{\vc{k}\}$ a local orthonormal basis 
\begin{equation} \label{eq:basis}
\{\vch{e}_p(\vch{k}),\vch{e}_c(\vch{k},\vch{\lambda}),\vch{e}_a(\vch{k},\vch{\lambda})\},
\end{equation}
where $\vch{e}_p$, corresponding to potential field, is collinear to $\vch{k}$, $\vch{e}_c$, corresponding to  compressible field, is perpendicular to $\vch{k}$ in the plane containing $\vch{k}$ and $\vch{\lambda}$ and $\vch{e}_a$, corresponding to Alfv\'enic field, is perpendicular to both $\vch{k}$ and $\vch{\lambda}$ \citep[see, for example,][]{Ch21}.

Then, we can write the following expression:
\begin{equation} \label{eq:proj}
T_{\alpha,ij}(\vch{k}) x_j = \hat{e}_{\alpha,i}(\vch{k}) \: x_j \hat{e}_{\alpha,j}(\vch{k}), 
\end{equation}
where $\alpha \in \{p,c,a\}$ and $x_i$ is an arbitrary vector.

\end{document}